\documentstyle[preprint,aps,amsfonts]{revtex}
\begin{document}


\newcommand\mathC{\mkern1mu\raise2.2pt\hbox{$\scriptscriptstyle|$}
                {\mkern-7mu\rm C}}		       
\newcommand{\mathR}{{\rm I\! R}}                


\begin{titlepage}

\hspace{8truecm} Newton Institute NI98001

\begin{center}
        {\large\bf Non-local properties of multi-particle density matrices}
\end{center}
\vspace{1 truecm}
\begin{center}
        N.~Linden${}^{a,}$\footnote{email: n.linden@newton.cam.ac.uk}, 
        S.~Popescu${}^{a,b,}$\footnote{email: s.popescu@newton.cam.ac.uk} and 
        A. Sudbery${}^{c,}$\footnote{email: as2@york.ac.uk}\\[0.4cm]
       ${}^a$ Isaac Newton Institute for Mathematical Sciences\\
        20 Clarkson Road\\
        Cambridge CB3 0EH\\ 
        United Kingdom\\
 \end{center}
\bigskip
\begin{center}
        ${}^b$BRIMS Hewlett-Packard Labs.\\
        Filton Road, Stoke Gifford\\
        Bristol, BS12 6QZ\\
        United Kingdom\\
\end{center} 
\bigskip
\begin{center}
        ${}^c$Department of Mathematics.\\
        University of York\\
        Heslington
        York, YO1 5DD\\
        United Kingdom\\
\end{center}\medskip 

\begin{center} January 1998\end{center}
 
\begin{abstract}
As far as entanglement is concerned, two density matrices of
$n$ particles are equivalent if they are on the same orbit of the group
of local unitary transformations, $U(d_1)\times\cdots\times U(d_n)$ (where 
the Hilbert space of particle $r$ has dimension $d_r$).  We show that for $n$ 
greater than 
or
equal to two, the number of independent parameters needed to
specify an $n$-particle density matrix up to equivalence is 
$\Pi_r d_r^2 - \sum_r d_r^2 + n - 1$. For $n$ spin-${1\over 2}$ particles we 
also show how to characterise generic orbits, both by giving an explicit 
parametrisation of the orbits and by finding a finite set of polynomial 
invariants 
which separate the orbits.    
\end{abstract}
 
\end{titlepage}
In this paper we take some further steps towards understanding multi-particle
entanglement by analysing the non-local properties of density
matrices of $n$ particles.  This continues the programme, begun in
\cite{LindenPopescu97a}, in which we gave a framework for studying the space of
pure states of $n$ spin-1/2 particles.  As discussed in \cite{LindenPopescu97a},
the space of pure states of $n$ spin-1/2 particles is  
$\mathC^{2^n}=\mathC^2\otimes...\otimes \mathC^2$; however not all the $2^n$
complex parameters have non-local significance: the group of local
transformations, $U(2)^n$ acts on the space of states and two states which may
be reached from each other by local actions are equivalent as far as their
non-local properties are concerned.  Each equivalence class of locally
equivalent density matrices is an {\em orbit} of this group. For many purposes, 
only parameters
describing non-local properties are significant; an example is that any good
measure of entanglement must be invariant under local transformations, and thus
it should be a function of non-local parameters only (here and henceforth we 
will
refer to parameters which are invariant under local transformations as
invariants).  A key question is to identify the invariants.

In this paper we will focus on density matrices and show that for $n\geq 2$, of
the $2^{2n}-1$ real parameters describing density matrices of $n$ spin-1/2
particles $2^{2n}-3n-1$ are invariant under local transformations, $U(2)^n$.  
This generalises to an arbitrary set of $n$ particles as $\prod_r d_r^2 - \sum_r
d_r^2 + n - 1$ where $d_r$ is the dimension of the state space of the $r$th 
particle. For $n$ spin-${1\over 2}$ particles we 
also show how to characterise generic orbits, both by giving an explicit 
parametrisation of the orbits and by finding a finite set of polynomial 
invariants 
which separate the orbits.    Thus given two density matrices we can
compute explicitly whether they are on the same orbit or not.  Other authors
have also discussed the use of invariants in discussing 
entanglement\cite{SchlienzMahler96}\cite{Grassletal97}
and applied invariant theory to quantum codes \cite{Rains97}.

In order to calculate the number of functionally independent invariants
it will be convenient to find the dimension of the orbit of a generic density
matrix under the group of local transformations.  The dimension of the orbit
is the number of parameters describing the location of a density matrix on the
orbit. The total number of parameters ($2^{2n}- 1$ 
real  parameters) describing the space of density matrices minus the number of
parameters  describing a generic orbit (the dimension of the orbit) gives the
number of  parameters describing the location of the orbit in the space of
orbits, i.e. the  number of parameters describing the non-local properties of
the density matrices.

To fix notation, it will be convenient to consider the case of a one-particle
density matrix first.  The space of pure states of a single spin-1/2 particle is
$\mathC^{2}$ and thus a density matrix is a $2\times 2$ complex matrix which is
hermitian, positive and with trace one, and may therefore be described by three 
real
parameters.  A particularly convenient representation of such a matrix is
\begin{equation}
\rho = {1\over 2}1_2 + \alpha_i\sigma_i,
\end{equation}
where $\alpha_i$, $i=1,2,3$ are real and
\begin{equation}
\sigma_1 = 
\left(\begin{array}{cc} 
0& 1 \\
1&0
\end{array}\right)
\quad
\sigma_2 = 
\left(\begin{array}{cc} 
0& -i \\
i&0
\end{array}\right)
\quad
\sigma_3 = 
\left(\begin{array}{cc} 
1& 0 \\
0&-1
\end{array}\right)
\quad
1_2 = 
\left(\begin{array}{cc} 
1& 0 \\
0&1
\end{array}\right)
.
\end{equation}

We note that
\begin{equation}
\sigma_i\sigma_j =i\epsilon_{ijk}\sigma_k +\delta_{ij}.
\end{equation}
Under a local transformation by a unitary matrix $U$, $\rho$ is transformed as
\begin{equation}
\rho\mapsto U\rho U^\dagger. \label{rhotransform}
\end{equation}
The group $U(2)$ is isomorphic to $U(1)\times SU(2)$ where, physically, the
$U(1)$ is the phase transformation of a state, represented by a unitary
matrix $e^{i\phi}1_2$. This element clearly leaves any density matrix invariant 
under the
transformation (\ref{rhotransform}) so that when considering the action 
(\ref{rhotransform}) we may restrict attention to elements of $SU(2)$.  In order 
to
find the number of invariants it will be more convenient to find the dimension
of a generic orbit under the action of $SU(2)$.  To do so one may work
infinitesimally. Thus,  associated to the action of the  Lie algebra of the 
group
of local transformations acting on the space of density matrices there is a
vector field: if we take an  element $T$ of a basis for the Lie algebra,  the 
action
of the group element $k=\exp i\epsilon T\in K$ on an element $\rho$ induces an
action on functions from $\rho$ to $\mathC$; and the vector field, $X_T$, 
associated to the Lie algebra element $T$ is found by differentiating:

\begin{equation}
X_T f(v) {\buildrel {\rm def}\over =} {\partial\over\partial\epsilon}
f(e^{i\epsilon T}\rho)\vert_{\epsilon =0} = 
{\partial\over\partial\epsilon}
f(\rho+\delta\rho)\vert_{\epsilon =0}.
\label{field}
\end{equation}
The linear span of vector fields at the point $\rho$ associated with the whole
Lie algebra forms the tangent space to the orbit at the point $\rho$ and so the
number of linearly independent vector fields at this point gives the dimension
of the orbit.

A general element of the Lie algebra in the spin-1/2 representation is given by
\begin{equation}
T=\eta_i\sigma_i
\end{equation}
and its action on the density matrix is to give an infinitesimal transformation
\begin{equation}
\delta\rho=i[T,\rho]
\end{equation}
where $[,]$ is the matrix commutator.

We may therefore calculate the three vector fields $X_1,X_2$ and $X_3$
associated to the Lie algebra elements $\sigma_1,\sigma_2$ and $\sigma_3$
as
\begin{equation}
X_1=
\alpha_2{\partial\over\partial\alpha_3}-\alpha_3{\partial\over\partial\alpha_2},
\quad 
X_2=
\alpha_3{\partial\over\partial\alpha_1}-
\alpha_1{\partial\over\partial\alpha_3},\quad
X_3=\alpha_1{\partial\over\partial\alpha_2}-\alpha_2{\partial\over\partial
\alpha_1}.
\label{vectorfields1} 
\end{equation}
We note that at generic values of $\alpha_1,\alpha_2,\alpha_3$ only two of these
vector fields are linearly independent since
\begin{equation}
\alpha_1 X_1 + \alpha_2 X_2 + \alpha_3 X_3 =0.
\end{equation}
Thus the dimension of the generic orbit is two and therefore of the three
parameters describing a generic density matrix, two are non-invariant leaving 
only one
invariant parameter, as one expects since only the single independent
eigenvalue of $\rho$ is invariant under local transformations.

We note that the effect of the transformations (\ref{rhotransform}) is to
act on  the vector ${\bf \alpha}$ by rotation by an orthogonal matrix, i.e. an
element of $SO(3)$ - this follows from the fact that $\alpha_i\sigma_i$ is the
representative of a Lie algebra element and the conjugation action 
(\ref{rhotransform}) is the adjoint action of the group on its Lie algebra. 
We may thus find a way of exhibiting the invariant under local transformations:
\begin{equation}
I=\alpha_i\alpha_j\delta_{ij} = |\alpha|^2
\end{equation}
where we have used the fact that $SO(3)$ has an invariant tensor
$\delta^{ij}$.  We note that this invariant may also be expressed as
\begin{equation}
I={\rm Tr}(\rho^2) - {1\over 2}.
\end{equation}
We now turn to the case of two-particle density matrices.  Such a density
matrix has 15 real parameters, and the maximum dimension that a generic orbit
could have is 6 (corresponding to two copies of $SU(2)$) if all the vector
fields corresponding to a basis of the Lie algebra were independent.  We will
show that the vector fields do indeed span 6 dimensions, and thus that there
are 9 non-local parameters.

We may write a density matrix as
\begin{equation}
\rho = {1\over 4}1_2\otimes 1_2 + \alpha_i\sigma_i\otimes 1_2 + \beta_i 1_2
\otimes \sigma_i + R_{ij}\sigma_i\otimes\sigma_j.\label{twoparticlerho} 
\end{equation}
The action of a Lie algebra element of the subgroup $SU(2)$ acting on the
first component of the tensor product is
\begin{eqnarray}
\delta^{(1)}\rho &=& i[\eta_k\sigma_k\otimes 1_2 , \rho]\nonumber\\
&=&\alpha_k\eta_m\epsilon_{mki}\sigma_i\otimes 1_2 + 
R_{kj}\eta_m\epsilon_{mki}\sigma_i\otimes\sigma_j,
\end{eqnarray} 
and that corresponding to a Lie algebra element of the subgroup $SU(2)$ acting 
on the
second component of the tensor product,
\begin{eqnarray}
\delta^{(2)}\rho &=& i[\eta_k 1_2\otimes\sigma_k  , \rho]\nonumber\\
&=&\beta_k\eta_m\epsilon_{mki}1_2\otimes\sigma_i + 
R_{ik}\eta_m\epsilon_{mkj}\sigma_i\otimes\sigma_j.
\end{eqnarray}
The vector fields corresponding to the six basis elements $\sigma_i\otimes 1_2$, 
$1_2 \otimes \sigma_i$ are 
\begin{eqnarray}
  X_k=\epsilon_{kim} \left( \alpha_i{\partial\over\partial \alpha_m} 
  + R_{ij}{\partial\over \partial R_{mj}}\right), \nonumber\\
 Y_k=\epsilon_{kim} \left(\beta_i{\partial\over\partial \beta_m}+ 
     R_{ji}{\partial\over\partial R_{jm}}\right) \label{vectorfields2}
\end{eqnarray}
     
Consider the set $X_i$ first: one can see that these three are linearly
independent at generic points by considering the coefficients of 
$\partial /\partial \alpha_i$, since a linear relation would have to be of the 
form 
$ \alpha_k X_k = 0$, but one can see that this relation will not hold for 
non-zero $\alpha$'s
by looking at the coefficients of the partial derivatives with respect to
$R_{ij}$.  Similarly by considering the coefficients of the partial derivatives
with respect to $\beta_1,\beta_2,\beta_3$, one sees that $Y_1,Y_2,Y_3$ are
linearly independent.  Finally, we note that the coefficients of
the partial derivatives with respect to
$\beta_1,\beta_2,\beta_3$ are zero for $X_1,X_2,X_3$ and the coefficients of
the partial derivatives with respect to
$\alpha_1,\alpha_2,\alpha_3$ are zero for $Y_1,Y_2,Y_3$ so that there can be no
linear relation at all between the six vector fields 
$X_1,X_2,X_3,Y_1,Y_2,Y_3$.
Thus the dimension of the orbit of a generic density matrix is 6 and thus the
number of non-local parameters, $15-6=9$.

In general, we can consider a system of $n$ particles with individual state 
spaces of dimensions $d_1, \ldots, d_n$. The density matrix is a hermitian $D 
\times D$ matrix 
with trace 1, where $D = d_1 d_2\ldots d_n$, and therefore requires $D^2 - 1$ 
real 
parameters which can be taken to be the coefficients $\alpha^{(1)},\ldots , 
\alpha^{(n)}, \ldots , R$ in an expansion
\begin{equation}
 \rho = {1\over D} 1_{d_1} \otimes \cdots 1_{d_n} + \sum_{r=1}^n 
\alpha_{i_r}^{(r)} 1 \otimes \cdots \otimes T_{i_r}^{(r)} \otimes \cdots \otimes 
1 
+ \cdots + R_{i_1\ldots i_n} T_{i_1}^{(1)}\otimes \cdots \otimes T_{i_n}^{(n)}
\end{equation}
where $T_{i_r}^{(r)}$ ($i_r = 1, \ldots, d_r^2 - 1$) are a basis set of 
traceless hermitian $d_r \times d_r$ matrices (generators of $SU(d_r)$). The 
action of an infinitesimal generator of $SU(d_r)$ acting on the $r$th factor of 
the tensor product is 

\begin{equation}
\delta^{(r)}\rho = c^{(r)}_{ijk}\eta_i (\alpha_j^{(r)}1 
\otimes \cdots \otimes T_k^{(r)} \otimes 
\cdots \otimes 1 + \cdots )\qquad (i,j,k = 1, \ldots , d_r^2 - 1)
\end{equation}
where $c_{ijk}^{(r)}$ are the structure constants of $SU(d_r)$. 
Thus the infinitesimal action of local transformations is given by a set of 
vector fields

\begin{equation}
X^{(r)}_i = c^{(r)}_{ijk}\left(\alpha^{(r)}_j {\partial \over \partial 
\alpha_k^{(r)}} 
+ \cdots +
R_{i_1\ldots j \ldots i_n}{\partial\over\partial R_{i_1\ldots k \ldots i_n}}
\right).
\end{equation}

Similar considerations to those used above for the case of two spin-${1\over 2}$ 
particles show that these vector fields are generically all 
independent. Thus the generic orbit has 
dimension $d_1^2 + \cdots d_n^2 - n$.  Since the space of density matrices has 
dimension $d_1^2\ldots d_n^2$, there are a total of 
\begin{equation} 
\prod_r
 d_r^2 - \sum_r d_r^2 + n - 1
\end{equation}
non-local invariants.

Let us now return to the case of $n\geq 2$ spin-1/2 particles and explicitly
identify a set of invariant parameters which characterise generic orbits.
To be explicit, consider the
case of three spin-1/2 particles with density matrix which may be written
as 
\begin{eqnarray}
\rho = {1\over 8}1_2\otimes 1_2\otimes 1_2 &+& 
\alpha_i\sigma_i\otimes 1_2\otimes 1_2 + 
\beta_i 1_2 \otimes \sigma_i \otimes 1_2
+\gamma_i 1_2 \otimes 1_2\otimes \sigma_i \nonumber\\
&+& R_{ij}\sigma_i\otimes\sigma_j\otimes 1_2
+ S_{ij}\sigma_i \otimes 1_2\otimes\sigma_j
+ T_{ij} 1_2\otimes\sigma_i\otimes\sigma_j\nonumber\\
&+& Q_{ijk} \sigma_i\otimes\sigma_j \otimes\sigma_k. 
\end{eqnarray}
The action 
by
a local unitary transformation on the first component in the tensor product 
induces the following transformations on the components of $\rho$
\begin{equation}
\alpha_i \mapsto L_{ij}\alpha_j; \quad
R_{ij} \mapsto L_{ik} R_{kj};\quad S_{ij} \mapsto L_{ik} S_{kj};\quad Q_{ijk} 
\mapsto L_{im} Q_{mjk}\end{equation}
where $  L_{ij}$ is an orthogonal matrix, and the other components of
$\rho$ do not change.  Similarly actions by a local transformations
on the second and third components of the tensor product induce
\begin{equation}
\beta_i \mapsto M_{ij}\beta_j; \quad
R_{ij} \mapsto M_{jk} R_{ik};\quad T_{ij} \mapsto M_{ik} T_{kj};\quad Q_{ijk} 
\mapsto M_{jm} Q_{imk}
\end{equation}
and
\begin{equation}
\gamma_i \mapsto N_{ij}\gamma_j; \quad
S_{ij} \mapsto N_{jk} S_{ik};\quad T_{ij} \mapsto N_{jk} T_{ik};\quad Q_{ijk} 
\mapsto N_{km} Q_{ijm}
\end{equation}
respectively, where $M_{ij}$ and $N_{ij}$  are orthogonal matrices independent 
of 
$L$.  

We max fix a canonical point on a generic orbit as follows: firstly let us
define
\begin{equation}
X_{ii^\prime} = Q_{ijk}Q_{i^\prime jk};\quad 
Y_{jj^\prime} = Q_{ijk}Q_{ij^\prime k};\quad
Z_{kk^\prime} = Q_{ijk}Q_{ijk^\prime},
\end{equation}
and perform unitary transformations on particles 1, 2 and 3 so as to
move to a point on the orbit in which $X,\ Y$ and $Z$ are diagonal;
generically the diagonal entries are distinct and we can arrange them 
in decreasing order ($X,\ Y$ and $Z$ are hermitian, positive
matrices).  The only remaining transformations which leave 
$X,\ Y$ and $Z$ in these forms are local unitary transformations which
induce orthogonal transformations in which $L_{ij}$, $M_{ij}$ and $N_{ij}$ are  
one of the
matrices
\begin{equation}
\left(\begin{array}{ccc} 
1& 0 &0\\
0&-1&0\\
0&0&-1
\end{array}\right)
\quad
\left(\begin{array}{ccc} 
-1& 0 &0\\
0&1&0\\
0&0&-1
\end{array}\right)
\quad
\left(\begin{array}{ccc} 
-1& 0 &0\\
0&-1&0\\
0&0&1
\end{array}\right)
\end{equation}
We may specify a canonical point on the generic orbit uniquely by
specifying that all the components of $\alpha$ have the same sign, and similarly 
for $\beta$ and 
$\gamma$.   This method works as long as $X,\ Y$ and $Z$
have distinct eigenvalues
and the components of $\alpha$, $\beta$ and $\gamma$ are not zero at
the canonical point on the orbit.
The parameters which describe
the generic orbits are the components of $\alpha,\beta,\gamma,R,S,T$ and $Q$
at the canonical point on the orbit.  We note that the number of parameters
describing the canonical point are the $2^6-1=63$ components of 
$\alpha,\beta,\gamma,R,S,T$ and $Q$ minus the $3\times 3=9$ constraints that
the non-diagonal elements of $X,\ Y$ and $Z$ are zero; thus the number
of non-local parameters is $54$ as given by the general formula.

We note that the fact that the canonical point, as constructed, is unique
means that all points on the same orbit will have the same canonical 
representative: conversely, if two density matrices $\rho_1$ and $\rho_2$
have the same canonical form, then 
\begin{equation}
U_1\rho_1 U_1^\dagger = \rho_{\rm canonical} = U_2\rho_2 U_2^\dagger
\end{equation}
for some $U_1$ and $U_2$, so that 
\begin{equation}
\rho_2 = (U_2^\dagger U_1)\rho_1 (U_2^\dagger U_1)^\dagger  
\end{equation}
and thus $\rho_1$ and $\rho_2$ are on the same orbit.

We now describe a {\it finite} set
of polynomial invariants which separate generic orbits by finding a 
set which allows one to calculate the components of
$\alpha,\beta,\gamma,R,S,T$ and $Q$ at this canonical point.  The complete 
infinite
set of polynomial invariants is found by contracting the indices of 
$\alpha,\beta,\gamma,R,S,T$ and $Q$ with the invariant tensors $\delta_{ij}$
and $\epsilon_{ijk}$.  However we may find a { finite} set of
invariants which separates generic orbits.
Firstly we note that ${\rm tr}(X),\ {\rm tr}(X^2)$ and ${\rm tr}(X^3)$ determine 
the
diagonal elements $\lambda_1^2,\ \lambda_2^2$ and $\lambda_3^2$ of $X$, and 
similarly for $Y$ and $Z$.  Now consider the three invariants
$A_{2n}=\alpha^T X^{n-1} \alpha$, $n=1,2,3$. We may write these three invariants 
in 
the following way:
\begin{equation}
\left(\begin{array}{ccc} 
1&1&1\\
\lambda_1^2&\lambda_2^2&\lambda_3^2\\
\lambda_1^4&\lambda_2^4&\lambda_3^4
\end{array}\right)
\left(\begin{array}{c} 
a_1^2\\
a_2^2\\
a_3^2
\end{array}\right)
=
\left(\begin{array}{c} 
A_2\\
A_4\\
A_6
\end{array}\right),
\end{equation}
where $a_1,\ a_2$ and $a_3$ are the components of $\alpha$ at the canonical 
point on the orbit.  The Vandermonde matrix
\begin{equation}
\Lambda=\left(\begin{array}{ccc} 
1&1&1\\
\lambda_1^2&\lambda_2^2&\lambda_3^2\\
\lambda_1^4&\lambda_2^4&\lambda_3^4
\end{array}\right)
\end{equation}
has  determinant 
$(\lambda_1^2-\lambda_2^2)(\lambda_2^2-\lambda_3^2)(\lambda_3^2-\lambda_1^2)$,
and we may solve for  $a_1^2,\ a_2^2$ and $a_3^2$ as long as $\det \Lambda$ is 
non-zero.  Also if the invariant
\begin{equation}
A_9=\epsilon_{ijk} \alpha_i (X\alpha)_j (X^2\alpha)_k = 
\alpha.(X\alpha)\wedge (X^2\alpha)= a_1a_2 a_3\det \Lambda
\end{equation}
is non-zero, then we may determine the sign of the  components of $\alpha$; 
recall that, by definition, all the components of $\alpha$ have the same sign at 
the canonical point.
The analogous
expressions $B_9, C_9$ determine the values of $\beta$ and $\gamma$ at the 
canonical 
point.  The values of the components of $R$ at the canonical point
may be calculated from the following nine invariants:
\begin{equation}
I_{r,s} = (X^{r-1}\alpha)_i (Y^{s-1}\beta)_j R_{ij}.
\end{equation}
These nine equations may be put together into a matrix form 
\begin{equation}
I = \left( (\Lambda F) \otimes (M G)\right) R,
\end{equation}
where $I$ and $R$ are  column vectors with nine components and the 
matrices $\Lambda,\ F,\ M$ and $G$ are
\begin{equation}
\Lambda = \left(\begin{array}{ccc} 
1&1&1\\
\lambda_1^2&\lambda_2^2&\lambda_3^2\\
\lambda_1^4&\lambda_2^4&\lambda_3^4
\end{array}\right);\quad
F = \left(\begin{array}{ccc} 
a_1&0&0\\
0&a_2&0\\
0&0&a_3
\end{array}\right);
\quad
M = \left(\begin{array}{ccc} 
1&1&1\\
\mu_1^2&\mu_2^2&\mu_3^2\\
\mu_1^4&\mu_2^4&\mu_3^4
\end{array}\right);\quad
G=
\left(\begin{array}{ccc} 
b_1&0&0\\
0&b_2&0\\
0&0&b_3
\end{array}\right),
\end{equation}
where $\mu_1^2,\ \mu_2^2$ and $\mu_3^2$ are the diagonal elements of $Y$. 
We note that det$(\Lambda F) = A_9$ and det$(MG)=B_9$, so since we
are assuming that these are non-zero we may invert the matrix
 equation to find the components $R_{ij}$.  The components of $S$ and $T$ may be 
found in a similar way.  Finally we may use the 27 invariants
\begin{equation}
I_{r,s,t} = (X^{r-1}\alpha)_i (Y^{s-1}\beta)_j (Z^{t-1}\gamma)_k Q_{ijk}.
\end{equation}
to find the components of $Q$ at the canonical point on the orbit in
terms of the $I_{r,s,t}$ (there will, of course, be some relations between
these components due the constraints that $X,\ Y$ and $Z$ are diagonal).

Thus, by showing that the following set of polynomial invariants is 
sufficient to
calculate the components of a generic density matrix at the canonical
point we have demonstrated that they characterise generic orbits:

\begin{eqnarray}
&{\rm tr}X^r, \quad {\rm tr}Y^r, \quad {\rm tr}Z^r &\nonumber\\
&\alpha^{\rm T} X^{r-1}\alpha, \quad \beta^{\rm T} Y^{r-1}\beta, \quad
\gamma^{\rm T} Z^{r-1}\gamma;&\nonumber\\
 &  \alpha.(X\alpha)\wedge (X^2\alpha),\quad 
  \beta.(Y\beta)\wedge (Y^2\beta),\quad
 \gamma.(Z\gamma)\wedge (Z^2\gamma)&\nonumber\\
& (X^{r-1}\alpha)_i(Y^{s-1}\beta)_jR_{ij},\quad
  (Y^{r-1}\beta)_i(Z^{s-1}\gamma)_jT_{ij},\quad
 (X^{s-1}\alpha)_i (Z^{r-1}\gamma)_j S_{ij};&\nonumber\\
& (X^{r-1}\alpha)_i(Y^{s-1}\beta)_j(Z^{t-1}\gamma)_kQ_{ijk};&
\end{eqnarray}
the indices $r,s,t$ range over the values $1,2,3$.

If two density matrices have different values of any of these invariants
they are not on the same orbit; if they have same value of all of these
invariants, and if $A_9$, $B_9$ and $C_9$ are non-zero, then the density
matrices are locally equivalent.

We note that the number of independent components of a generic density
matrix at the canonical point is equal to the nubmer of functionally
independent parameters calculated at the beginning of this letter.
However, the number of polynomial invariants needed to characterise the
generic orbit is greater than this; this is related to the fact that the
ring of invariants is non-polynomial, i.e. that the geometry of the
space of orbits is non-trivial.

The procedure given above can be used for all $n\geq 2$: use the tensors of
highest rank and rank one in the expression for $\rho$ to fix a canonical point 
on the orbit; the polynomials which separate the generic orbits
are the analogues of those used in the case $n=3$.

In the case of $n=2$ this method can be used but there is some redundancy in the 
description we have given: the matrices $X_{ii^\prime} = R_{ij}R_{i^\prime j}$ 
and $Y_{jj^\prime} = R_{ij}R_{i j^\prime}$ (using the notation of 
(\ref{twoparticlerho}))
have the same eigenvalues and the matrix $R_{ij}$ is diagonal
at the canonical point.  In this case there are nine functionally independent
invariants which specify the squares of the non-zero components of $\alpha$ , 
$\beta$ and $R$ at the canonical point on a generic orbit: ${\rm tr}X^n$, 
$\alpha^T X^{m-1} \alpha$ and 
$\beta^T Y^{p-1}\beta$, where $n,m,p$ take the values $1,2,3$.  
Additional invariants are needed to specify the signs of the non-zero 
components.  
The five invariants $\alpha.(X\alpha)\wedge (X^2\alpha)$, 
$\beta.(Y\beta)\wedge (Y^2\beta)$ and $\alpha
X^{r-1}R\beta, 
\quad r=1,2,3$, are sufficient to determine these signs for generic orbits and 
hence 
separate these orbits. In fact, using slightly different arguments, one can show 
that, in this case, one can reduce the number of polynomial invariants to ten, 
namely
tr$X$, tr$X^2$, det$R$, $\alpha^T X^{r-1}\alpha$, $\alpha^T X^{r-1}R
\beta$, $r=1,2,3$ and $A_9$, which are subject to a single relation
expressing $A_9^2$ as a function of the other invariants.

The general idea of investigating canonical points on orbits in the way we have 
described is also appropriate for higher spins, but the situation is 
somewhat more complicated.  Consider the example of two particles of spin one in 
which case the unitary group under which $\rho$ transforms is $SU(3)$.  $\rho$ 
may be written as
\begin{equation}
\rho = {1\over 64} 1_8\otimes 1_8 + \alpha_i T_i\otimes 1_8 + 
\beta_i 1_8\otimes T_i + R_{ij} T_i\otimes T_j
\end{equation}
where $T_i,\ i=1...8$ are representatives of a basis for the Lie algebra of 
$SU(3)$ in the adjoint representation and $1_8$ is the $8\times 8$ identity 
matrix.  However, the adjoint representation  of $SU(3)$ is equivalent
not to  
$SO(8)$ but to an eight dimensional subgroup of it; this means that we 
cannot transform 
$\rho$ so that $RR^T$ and $R^TR$ are diagonal so the canonical form is 
rather more complicated than in the case of spin-1/2 particles.

In summary we have shown how to calculate the number of functionally 
independent parameters needed to determine whether or not two density matrices 
are
locally equivalent.  We have also shown how to characterise the generic classes 
of locally 
equivalent density matrices
of $n$ spin-1/2 particles by two methods: (a) by finding an explicit set of 
non-polynomial 
invariants
(the components of the density matrices at the  canonical points on the orbits) 
and (b) by
finding an explicit finite set of polynomial invariants.  These methods work for 
generic density matrices;  
in a future publication
we intend to give a systematic method for characterising classes of locally 
equivalent 
non-generic 
density matrices.  In particular this will give a basis for the ring of 
invariants.  We 
note
that the canonical point on certain types of non-generic orbit has
non-trivial stability group; this is a signature
that density matrices on this orbit have special types of entanglement
\cite{LindenPopescu97a}.

 \bigskip 
\noindent{\large\bf Acknowledgments}

\noindent
We are 
very grateful to the Leverhulme and Newton Trusts for the financial support
given to NL.

\end{document}